\documentclass[aps,prb,twocolumn,superscriptaddress,showpacs,final,letterpaper]{revtex4}

\usepackage{amsmath}
\usepackage{amssymb}
\usepackage{graphicx}
\usepackage{dcolumn}
\usepackage{bm}

\begin{document}

\newcommand{\NaA}{Na$_{2}$AC$_{60}$}
\newcommand{\NaRb}{Na$_{2}$RbC$_{60}$}
\newcommand{\NaK}{Na$_{2}$KC$_{60}$}
\newcommand{\NaCs}{Na$_{2}$CsC$_{60}$}
\newcommand{\Rb}{Rb$_{3}$C$_{60}$}
\newcommand{\K}{K$_{3}$C$_{60}$}
\newcommand{\A}{A$_{3}$C$_{60}$}
\newcommand{\C}{C$_{60}$}
\newcommand{\Carbon}{$^{13}$C}
\newcommand{\Na}{$^{23}$Na}
\newcommand{\Pa}{$Pa\bar{3}$}
\newcommand{\Fm}{$Fm\bar{3}m$}
\newcommand{\tp}{T$^\prime$}

\title{Influence of local fullerene orientation on the electronic properties of A$\boldsymbol{_{3}}$C$\boldsymbol{_{60}}$ compounds}

\author{P{\'e}ter~\surname{Matus}}
\email[E-mail: ]{matus@szfki.hu}
\affiliation{Laboratoire de Physique des Solides, UMR~8052, Universit{\'e} Paris-Sud, B{\^a}timent~510, 91405 Orsay cedex, France}
\affiliation{Research Institute for Solid State Physics and Optics of the Hungarian Academy of Sciences, P.O.B.~49, H-1525 Budapest, Hungary}

\author{Henri~\surname{Alloul}}
\affiliation{Laboratoire de Physique des Solides, UMR~8052, Universit{\'e} Paris-Sud, B{\^a}timent~510, 91405 Orsay cedex, France}

\author{Gy{\"o}rgy~\surname{Kriza}}
\affiliation{Research Institute for Solid State Physics and Optics of the Hungarian Academy of Sciences, P.O.B.~49, H-1525 Budapest, Hungary}

\author{V{\'e}ronique~\surname{Brouet}}
\affiliation{Laboratoire de Physique des Solides, UMR 8052, Universit{\'e} Paris-Sud, B{\^a}timent 510, 91405 Orsay cedex, France}

\author{Philip~M. \surname{Singer}}
\altaffiliation[Present address:~]{Schlumberger-Doll Research, 36 Old Quarry Road,
Ridgefield, CT 06877, USA}
\affiliation{Laboratoire de Physique des Solides, UMR 8052, Universit{\'e} Paris-Sud, B{\^a}timent 510, 91405 Orsay cedex, France}

\author{Slaven~\surname{Garaj}}
\altaffiliation[Present address:~]{Department of Physics, Harvard University, Cambridge MA 02138, USA}
\author{L{\'a}szl{\'o}~\surname{Forr{\'o}}}
\affiliation{Laboratoire de Nanostructures et Nouveaux Mat{\'e}riaux Electroniques,
Institute of Physics of Complex Matter, School of Basic Sciences at the Ecole Polytechnique F{\'e}d{\'e}rale de Lausanne, 1015 Lausanne, Switzerland}
\date{\today}

\begin{abstract}

We have investigated sodium containing fullerene superconductors \NaA{}, A = Cs, Rb, and K, by \Na{} nuclear magnetic resonance (NMR) spectroscopy at 7.5~T in the temperature range of 10 to 400~K. Despite the structural differences from the \Rb{} class of fullerene superconductors, in these compounds the NMR line of the tetrahedrally coordinated alkali nuclei also splits into two lines (T and \tp{}) at low temperature. In \NaCs{} the splitting occurs at 170~K; in the quenched cubic phase of \NaRb{} and \NaK{} we observe split lines at 80~K. By detailed investigations of the spectrum, spin-spin and spin-lattice relaxations as well as spin-echo double resonance (SEDOR) in \NaCs{} we show that these two different tetrahedral sites are mixed on a microscopic scale. The T and \tp{} sites differ in the orientation of first-neighbor \C{} molecules. We present evidence that the orientations of neighboring molecules are uncorrelated. Thermally activated molecular reorientations cause an exchange between the T and \tp{} sites and motional narrowing at high temperature. We infer the same activation energy, 3300 K, in the temperature range 125 to 300~K. The spin-lattice relaxation rate is the same for T and \tp{} down to 125~K but different below. Both the spin-lattice relaxation rate and Knight shift are strongly temperature dependent in the whole range investigated. We interpret this temperature variation by the effect of phonon excitations involving the rigid librational motion of the \C{} molecules. By extending the understanding of the structure and molecular dynamics of \C{} superconductors, these results may help in clarifying the effects of the structure on the superconducting properties.
\end{abstract}

\maketitle

\section{Introduction}

Buckminsterfullerenes intercalated with alkali atoms represent, with no doubts, the most investigated class of \C{} materials. The main source of interest is the superconductivity with unusually high critical temperature in \A{} compounds (A stands for an alkali metal).\cite{hebard,tanigaki2,gunnarsson_review} The charge transfer from alkali atoms to fullerene molecules in these systems is virtually complete,\cite{fu92} and since the lowest unoccupied molecular orbital (LUMO) of \C{} is triply degenerate,\cite{haddon} several A$_n$\C{} compounds can be synthesized\cite{fischer} by varying A and n. Many of these are strongly correlated metals exhibiting unusual collective phenomena besides superconductivity such as transition to an insulating Mott--Jahn--Teller state.\cite{tosatti,forro,prassides_springer} 

Similarly to most \C{} crystals, A$_n$\C{} compounds are plastic crystals at high temperature, \emph{i.e.} the orientation of the \C{} balls, located on a face-centered cubic (fcc) lattice, varies fast in
time.\cite{tycko2} The alkali atoms are situated at interstices between the \C{} molecules.\cite{fischer} The relative orientation and the dynamic variation of the orientations of the fullerene molecules play a crucial role in the physical properties of alkali fulleride materials. Perhaps the most spectacular example is the formation of one-dimensional polymeric chains\cite{polymer} by [2 + 2] cycloadditional solid state reaction in certain A$_n$\C{} crystals. The rate limiting factor of the reaction is the fraction of the neighboring \C{} molecules with relative orientation appropriate for the cycloaddition to occur. At low temperatures, where the molecular orientations are frozen, a substantial amount of orientational disorder is retained.\cite{stephens,heiney2}

It has been realized early that the orientational order plays an important and nontrivial role also in the electronic and superconducting properties. A remarkable example is the different lattice constant dependence of the superconducting transition temperature in \Fm{} and \Pa{} superconductors with different orientational structure.\cite{yildirim} In a brilliant series of experiments Yang \emph{et al.}\cite{yang_arpes} and Brouet \emph{et al.}\cite{brouet_arpes1,brouet_arpes2} have investigated alkali fulleride monolayers deposited on different Ag substrates by angle resolved photoemission spectroscopy (ARPES). 
They have found different \C{} orientational structures with remarkably different electronic properties depending on what Ag crystalline plane the monolayers have been deposited on. 

For each \C{} molecule there is one octahedrally coordinated and two tetrahedrally coordinated voids in the fcc structure.\cite{stephens} 
In the \A{} binary and A$_2$A$\!^\prime$\C{} ternary compounds (A and A$\!^\prime$ indicate different alkali metals) all these voids are occupied by alkali ions. 
Several aspects of the structure depend on the alkali ionic radii. In the A$_2$A$\!^\prime$\C{} case, the larger alkali ions prefer the octahedral voids\cite{stenger} as these are more spacious than the tetrahedral voids. 
The energy associated with the electronic interactions between the \C{} molecules, as well as between the molecules and alkali ions is minimized in an orientationally ordered structure.\cite{launois} 
At high enough temperatures, however, the entropy associated with the rotational freedom of the molecules dominates over these electronic interactions. 
Here the orientational correlations are weak, the molecules can be described as spheres to good approximation, and an fcc structure with space group $Fm\bar{3}m$ forms. 
Upon decreasing the temperature, orientational order appears in a first order phase transition. 
In pristine \C{} this transition occurs at 263~K.\cite{heiney2,david} 

If the alkali ion occupying the tetrahedral voids is sodium, \NaA{}, with a radius small enough to fit into the tetrahedral voids, the same simple cubic (sc) orientational structure forms as in pure \C{} (space group: $Pa\bar{3}$).\cite{prassides} 
If, however, the ionic radius is larger than the size of the tetrahedral void (A = K, Rb, or Cs), the \A{} salts form an fcc structure with merohedral disorder in the fullerene orientations (space group: $Fm\bar{3}m$).\cite{stephens} 
One can conclude that in the first case interfullerene interactions, in the second case, the fullerene--alkali interactions will determine the orientational structure. 

In the $Fm\bar{3}m$ fcc structure (\emph{e.g.}, \Rb{}) the cubic axes are parallel to the twofold axes of the molecule bisecting the C--C double bonds. This is sometimes referred to as ``standard orientation''.\cite{stephens} 
The bonds, however, can take two different orientations, and these two standard orientations are occupied randomly and with equal probability in this fcc structure (merohedral disorder).\cite{stephens} 
In either case, alkali ions at tetrahedral voids face the centers of carbon hexagons leaving more room for the ions. 
In the $Pa\bar{3}$ sc structure (\emph{e.g.}, \NaA{}) the fullerene molecules are rotated away from the standard orientation about the [111] axes so that electron-rich double bonds face electron-poor regions on neighboring molecules in order to reach an energetically favorable state.\cite{david}

One of the oldest unresolved problems related to local \C{} orientational order in alkali fullerene salts is the so-called \tp{} problem in NMR experiments. 
In \A{} compounds the NMR line of the alkali nuclei in tetrahedrally coordinated voids is split into two lines at sufficiently low temperature (T--\tp{} splitting). 
The phenomenon has been discovered by Walstedt and coworkers\cite{walstedt} in \Rb{} below 370~K. 
They have found that the intensity ratios of the octahedral (O) and the two tetrahedral (T and \tp{}) NMR lines are close to O : T : \tp{} = 3 : 6 : 1. 
In addition they have shown by spin-echo double resonance (SEDOR) experiments that the \tp{} line originates from a tetragonal site slightly different from the site giving rise to the T line (hence the notation \tp{}). 
Since this result challenged the basic understanding of the structure of \A{} compounds, it has attracted much interest.\cite{stenger,pc1,pc2,mehring,yoshke,pennington_prb,pennington_prl,luders} 
Later this phenomenon has been demonstrated in the whole family of merohedrally disordered \A{} fullerides of fcc structure.\cite{stenger} 
Moreover, the splitting of the NMR 
line of the octahedral site (O--O$^\prime$ splitting) has also been observed\cite{mehring} by magic angle spinning (MAS).

In this paper we present a detailed NMR investigation of \NaA{} where A = K, Rb, or Cs. 
We have established the T--\tp{} splitting in all these compounds demonstrating that the splitting occurs not only in the fcc structure but also in the sc structure. 
To uncover the origin of the phenomenon, we have measured several NMR properties such as spectrum, spin-lattice and spin-spin relaxation time, spin-echo double resonance (SEDOR). 
We propose that the T--\tp{} splitting originates from two distinct orientational patterns of the neighboring \C{} molecules. 
The frequency of molecular reorientations of the molecules increases with increasing temperature leading to site exchange and ultimately motional narrowing of the two T sites. 

A preliminary encounter of some of these results has already been given in two brief conference proceedings articles.\cite{matus_amr,matus_jsc} In Ref.~\onlinecite{matus_amr} we present \Na{} spectra of \NaCs{} as well as the T--\tp{} splitting frequency and \Carbon{} line width. In Ref.~\onlinecite{matus_jsc} we demonstrate the SEDOR effect, and display $T_1$ and $T_2$ data. In the present article we insert these results in the context of a broad set of experiments and provide a unified analysis and discussion of all our NMR findings.

\section{Experimental}
The \NaCs{}, \NaRb{} and \NaK{} powder samples investigated in this study were prepared by conventional solid state reaction method\cite{prassides} using commercially available high-purity \C{} powder and alkali metals in soichiometric proportions. 
The samples were sealed in quartz tubes under He atmosphere. The phase purity was confirmed by x-ray diffraction. 
SQUID susceptibility measurements in \NaCs{} revealed a superconductivity transition temperature of 12~K in good agreement with earlier findings.\cite{tanigaki_nacs_supra} 

For the NMR measurements 50 to 100~mg powder was used in 5~mm diameter quartz tubes placed in a magnetic field $B_0$ = 7.5~T ($\nu_\text{Na}$ = 84.4~MHz). 
Data were collected by a home-built spectrometer. 
Spin-lattice and spin-spin relaxation measurements have been carried out with saturation recovery and spin-echo sequences, respectively. 
More details of each experiment are given in Section~\ref{results}.

\section{Results}
\label{results}
\begin{figure}
 \centering
  \includegraphics[width=0.75\columnwidth]{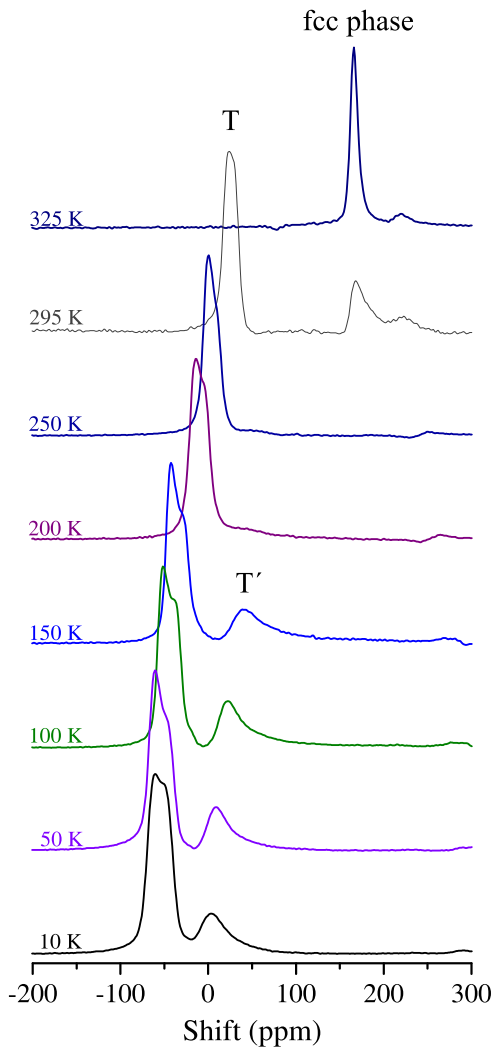}
  \caption{\label{spectrum} (color online) \Na{} NMR spectra of \NaCs{} in 7.5~T magnetic field. 
  The temperatures are indicated on the left of the spectra. 
  Shifts are measured from a NaI solution.} 
 \end{figure}

In \NaA{} (A = K, Rb, or Cs) salts the sodium ions occupy the tetrahedral voids in the structure,\cite{prassides} whereas the other alkali ion of larger size is situated in an octahedral void. 
The \Na{} NMR spectra of \NaCs{} taken at various temperatures are displayed in Fig.~\ref{spectrum}. 
At around 300~K one can observe an abrupt change in the NMR frequency what we 
associate with the first order fcc--sc phase transition.\cite{tanigaki,saito} 

Another remarkable feature of the spectra shown in Fig.~\ref{spectrum} is the appearance of a new line below about 170~K. 
We will discuss the properties of this line in details and show that this is essentially the same phenomenon as the T--\tp{} splitting in merohedrally disordered \A{} compounds, therefore, we denote this spectral component by \tp{}. 
The intensity of the \tp{} line increases with decreasing temperature reaching a spectral weight of $29\pm5$\% in the low-temperature limit. 

 \begin{figure}
 \centering
  \includegraphics[width=\columnwidth]{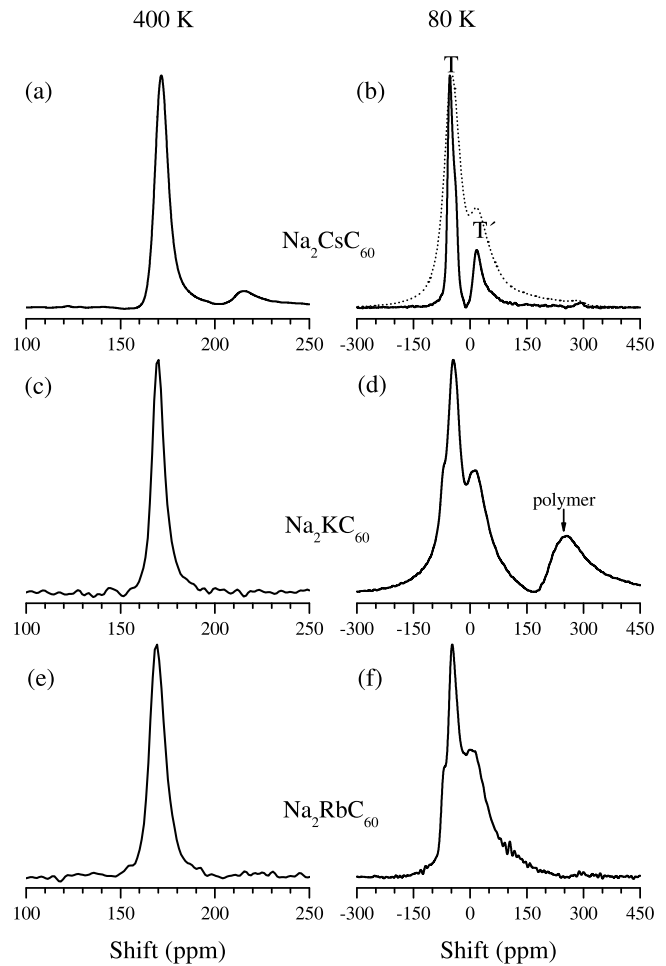}
  \caption{\label{Na2Aspect} \Na{} NMR spectra of \NaA{} compounds in 7.5~T magnetic field at two temperatures, 400~K (spectra on the left) and 80~K (spectra on the right). 
  (a) and (b)~\NaCs{}; (c) and (d)~\NaK{}; (e) and (f)~\NaRb{}. 
The dashed line in panel (b) is the convolution of the spectrum with a Gaussian of width 3~kHz (full width at half maximum).} 
 \end{figure}

The T--\tp{} splitting in $Pa\bar {3}$ fullerene structures is not restricted to \NaCs{}; we have detected the phenomenon in \NaRb{} as well as in \NaK{}. 
Both \NaRb{} (Refs.~\onlinecite{na2rb_polymer,na2rb_polymer2}) and \NaK{} (Ref.~\onlinecite{brouet_aip}) polymerize spontaneously when cooled slowly in the temperature range of 250~K to 220~K. 
Rapid cooling, however, largely suppresses the polymer formation.\cite{na2rb_polymer} 
In order to avoid the polymerization and preserve the cubic phase, we have adopted the following cooling protocol. 
The samples have been warmed up to 400~K in the NMR probe head, well into the fcc phase, then immersed into liquid nitrogen. 
The \Na{} NMR spectra of \NaRb{} and \NaK{} taken at 400~K in the fcc phase and 80~K in the quenched sc phase are compared in Fig.~\ref{Na2Aspect}. 
The spectra of \NaCs{} taken at the same temperatures are also included in the figure for comparison. 
At 80~K the T--\tp{} splitting is obvious in all the three compounds. 
A third line at about 240~ppm is also observed in \NaK{} at 80~K. 
Based on the \Na{} NMR spectrum of the polymer phase,\cite{brouet_aip} we identify this line as arising from a residual polymer phase in our quenched sample. 

Although the line widths of the three \NaA{} salts at high temperature are identical, at 80~K \NaRb{} and \NaK{} have larger line widths than \NaCs{}. 
The larger line widths may arise from a larger defect concentration in these compounds. 
The quenching process may increase the defect concentration, although in \NaCs{} we do not observe any cooling rate dependence of the line width in the cooling rate range of $10^{-3}$ to $10^2$~K/s. 
If we apply an additional 3-kHz Gaussian broadening to the 80-K spectrum of \NaCs{} [see 
the dashed line in Fig.~\ref{Na2Aspect}(b)] to mimic the larger line widths of the other 
two compounds, the three spectra become virtually identical. 
Albeit that all \NaA{} materials display the T--\tp{} splitting, in the following we restrict our detailed NMR investigations to \NaCs{} because of the absence of the complicating polymer phase and because the narrower lines allow a better resolution.

Finally we comment on the weak satellite line\cite{saito} at 215 ppm in the fcc phase of \NaCs{} [see Fig.~\ref{Na2Aspect}(a)]. 
This line appears below 580~K and disappears at the fcc--sc phase transition. 
Although intriguing it is, the fcc satellite line seems to be unrelated to the T--\tp{} splitting in the orientationally ordered phase as it is absent in the other two \NaA{} compounds. 

 \begin{figure}
 \centering
  \includegraphics[width=0.95\columnwidth]{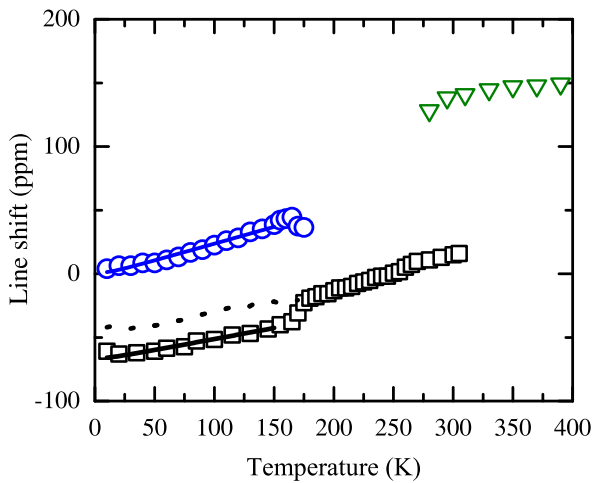}
  \caption{\label{shift} (color online) Temperature dependence of \Na{} NMR line shifts in \NaCs{}.  Squares and circles denote the T and \tp{} lines, respectively. The dotted line represents the first moments of the spectra and the triangles indicate the line positions in the fcc phase. The solid lines are fits of Eq.~(\ref{einsteinfit}).}
 \end{figure}

The temperature dependence of the first moment of the spectrum is shown in Fig.~\ref{shift}. 
The positions of the T and \tp{} lines are also indicated at low temperatures; here the first moment is the weighted average of the positions of the two lines with the corresponding spectral weights. 
At high temperature, the first moment is simply the position of the single tetrahedral line. 
It is remarkable that no singularity appears in the first moment at the temperature of the line splitting. 
Based on this observation the possibility that the \tp{} line originates from a segregated phase with different structure is very unlikely. 

As a more direct probe to exclude phase separation, T--\tp{} SEDOR experiments\cite{slichter} have also been performed. 
By the SEDOR technique one can probe if two species of nuclei, $\alpha$ and $\beta$, are spatially close to each other making use of the strong distance dependence of magnetic dipolar coupling between the nuclei. 
In the experiment the spin-echo of the $\alpha$ nuclei is measured with a $\pi /2-\tau -\pi$ pulse sequence under two different conditions: 
In one case a $\pi$ pulse at the resonant frequency of the $\beta$ nuclei is applied simultaneously with the $\pi$ pulse of the spin-echo sequence. 
In the other case the $\beta$ nuclei are not excited. 
Since the excitation of the $\beta$ nuclei influences the echo signal via the dipolar interaction between the $\alpha$ and $\beta$ nuclei, the difference between the two echo signals is a measure of the strength of dipolar coupling between them. 

To quantify the effect the so-called SEDOR fraction\cite{sedorfraction} SF is introduced with the following definition

\begin{equation}
\mathrm{SF} = 1 - \frac{I(\beta\; \text{on})}{I(\beta\; \text{off})}
\label{sf}	
\end{equation}

\noindent
where $I(\beta\; \text{on})$ and $I(\beta\; \text{off})$ are the spin-echo intensities with the excitation to the $\beta$ nuclei on and off, respectively. 
Maximal echo suppression gives $\text{SF} = 1$ while $\text{SF} = 0$ means no measurable dipolar coupling between the two species. 

 \begin{figure}
	\centering
      	\includegraphics[width=0.95\columnwidth,keepaspectratio]{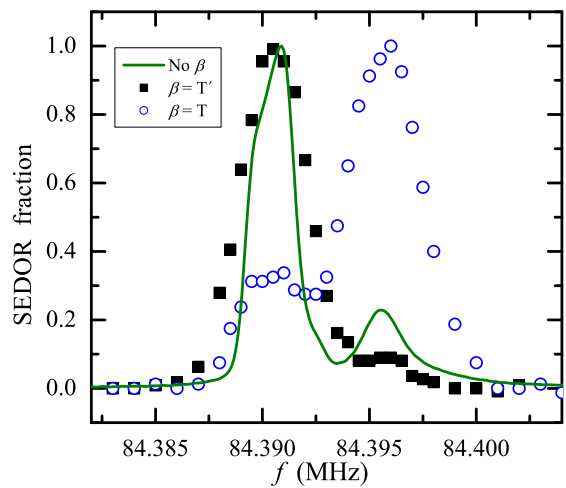}
      	\caption{\label{sedorsweep} (color online) SEDOR fractions as a function of the $f_\beta$ frequency at 80~K. The continuous line is the NMR spectrum for comparision. During the experiment the $f_\alpha$ frequency is kept constant at T (84.3905~MHz; squares) or \tp{} (84.3955~MHz; circles)}.
 \end{figure}

Figure~\ref{sedorsweep} demonstrates the SEDOR effect between the T and \tp{} lines. 
In these experiments the $\alpha$ nuclei were the T, and the \tp{} nuclei were 
the $\beta$ nuclei, and vice versa.\cite{matus_jsc}
Since both the $\alpha$ and $\beta$ nuclei are \Na{} and the splitting in frequency is small, it was difficult to achieve good selective excitation of the two lines. 
For such an excitation we used low-amplitude radiofrequency pulses with typical $\pi /2$ length of 150~$\mu$s. The interpulse delay was 5~ms, much longer than the $\pi /2$ length. 
The $\alpha$ frequency $f_\alpha$ was kept at the frequency of one of the T lines while the $\beta$ frequency $f_\beta$ was swept with small increments in frequency through the spectrum. 
For $f_\alpha \approx f_\beta$ the expectation is $\text{SF}\approx 1$. In the absence of dipolar interaction between the two species, $\text{SF} \approx 0$ is expected if $f_\beta$ is far from $f_\alpha$. However, a marked increase of the SEDOR fraction is observed in the experiments whenever $f_\beta$ is the frequency of the \emph{other} tetrahedral line. This effect indicates that the T and \tp{} species are strongly coupled, \emph{i.e.}, they are mixed on a microscopic scale. 

 \begin{figure}
	\centering
      	\includegraphics[width=\columnwidth]{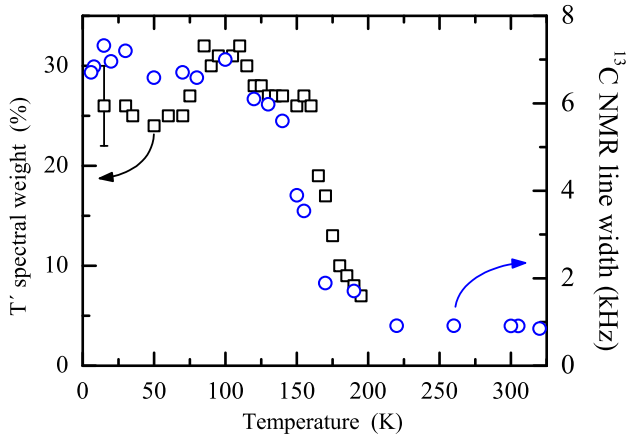}
      	\caption{\label{linewidth} (color online) \tp{} spectral weight (squares, left scale) together with the \Carbon{} NMR line width (circles, right scale) as a function of temperature.}  	
 \end{figure}
 
The intensity of the \tp{} line increases fast with decreasing temperature 
(Fig.~\ref{linewidth}) and below about 120~K it saturates within experimental error at $29 \pm5$\% of the total intensity of the T and \tp{} lines. 
The width of the \Carbon{} NMR line is also shown in Fig.~\ref{linewidth} to demonstrate the similar temperature dependence of the two quantities,\cite{matus_amr} just like in \A{} salt with fcc structure.\cite{mehring,yoshke}
Since the broadening of the \Carbon{} line signals the slowing down of molecular reorientations, this finding suggests that the splitting of the T line is related to fullerene molecular dynamics and orientational order. 

 \begin{figure}
  \includegraphics[width=0.95\columnwidth]{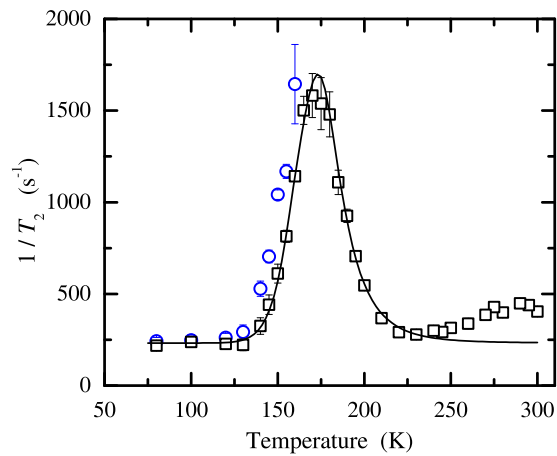}
  \caption{\label{t2} (color online) Temperature dependence of the \Na{} NMR spin-spin relaxation rate in \NaCs{}. The squares and circles represent the T and \tp{} lines, respectively. 
The solid line is a fit described in the text.}
 \end{figure} 
 
The temperature dependence of the spin-spin relaxation rate $1/T_2$ for both 
the T and \tp{} lines is displayed in Figure~\ref{t2}. The main feature is a peak at 
170~K, \emph{i.e.}, at the temperature of the line splitting.\cite{matus_jsc}
Otherwise $1/T_2$ displays a weak temperature dependence without special features. 
The spin-spin relaxation rate of the T$^\prime$ line increases sharply when the line splitting temperature is approached from below.

 \begin{figure}
  \includegraphics[width=0.95\columnwidth]{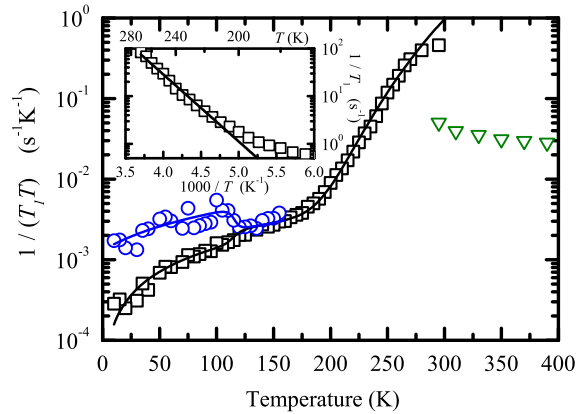}
  \caption{\label{t1} (color online) Temperature-normalized $^{23}$Na NMR spin-lattice relaxation rate $1/(T_1T)$ in \NaCs{} as a function of temperature, squares: T line; circles: \tp{} line; triangles: fcc phase. 
The inset is the $1/T_1$ as a function of inverse temperature. 
The solid lines are fits described in the text.}
 \end{figure} 

We are not aware of any earlier experiment in which the relaxation times of both the T and \tp{} lines were measured separately. 
The fact that the T--\tp{} splitting is well resolved in our \NaCs{} sample makes this compound an ideal candidate for these investigations. 
The temperature-normalized spin-lattice relaxation rate $1/(T_1 T)$ for both tetrahedral lines is displayed in Fig.~\ref{t1}. 
A characteristic feature is the bifurcation of the relaxation rates with decreasing temperature. This bifurcation occurs, however, at about 125~K, \emph{i.e.}, at a temperature much lower than the splitting of the spectrum line at 170~K. 
The separation of the two temperatures indicates the absence of a sharp transition and points to a gradual crossover. 

Although $1/(T_1T)$ of \Carbon{} and $^{133}$Cs nuclei\cite{brouet3} is constant below about 80~K as expected for a metal with Korringa relaxation, $1/(T_1T)$ of the \Na{} nuclei is strongly temperature dependent down to the lowest investigated 
temperature, 10~K. 
We offer an explanation based on the librational motion of the \C{} molecules in the Discussion. 
The rapid increase of sodium $1/(T_1T)$ in the high temperature range suggest a dominant relaxation channel governed by thermal activation in this temperature range, a feature similar to most \C{} compounds. 

\section{Discussion}
\label{discussion}

After a brief description of the orientational ordering transition in \NaCs{} in Section~A, we turn our attention to the T--\tp{} splitting. 
In Section~B we propose that the T and \tp{} sites originate from different \C{} orientational environments and put forward that site exchange due to fullerene molecular reorientations leads to the merger of these lines at high temperature. 
In Section~C we argue that the T and \tp{} sites differ in the orientation of first neighbor \C{} molecules.
Finally, in Section~D we analyze the anomalous low-temperature behavior of the Knight shift and spin-lattice relaxation rate. 

\subsection{The fcc--sc phase transition}
\label{ordering}

An fcc--sc phase transition in \NaCs{} at 299~K has been described by x-ray diffraction measurements\cite{prassides_xray} and found to be similar to the first order orientational ordering transition in \C{}. 
This transition has also been observed with sodium NMR by Saito \emph{et al.}\cite{saito} 
The discontinuous change in the NMR frequency in our measurements and the coexistence of the lines characteristic of the two phases in the 295-K spectrum (Fig.~\ref{spectrum}) are in agreement with a first order transition. 

In the high-temperature fcc phase the sodium ions sit at the center of the tetrahedron formed by the first-neighbor \C{} molecules. 
The site symmetry is cubic, the electric field gradient (EFG) tensor is zero, and there is no quadrupolar broadening of the NMR line. 
In contrast, in the low-temperature sc phase the site symmetry lowers to $C_3$, consequently the sodium ion moves out of the center of the tetrahedron along the threefold axis and uniaxial EFG develops. 
Indeed, the broadening of the NMR line is obvious below the transition temperature (see Fig.~\ref{spectrum}). 
The line shape is characteristic of the powder pattern of a $\frac{1}{2}\rightarrow\! -\frac{1}{2}$ central transition of a nucleus of spin $I=3/2$ under the influence of a uniaxial EFG. 
From an analysis of the line shape, we infer a quadrupolar splitting parameter of $\nu_Q =540 \pm 30$~kHz in the low-temperature limit. 
This is still smaller than the $680 \pm 20$~kHz measured in Na$_2$\C{} (Ref.~\onlinecite{brouet}) where the sodium site symmetry is identical to that of \NaCs{} indicating a smaller displacement of the sodium nuclei from the center of the tetrahedra in the latter compound. 

Although the 295-K phase transition in \NaCs{} was observed in previous sodium NMR studies by Saito \emph{et al.},\cite{saito} the T--\tp{} splitting was not detected in these experiments. 
The width of the T line they detected is larger than the frequency difference of the T and \tp{} lines. 
The lower applied magnetic field may be responsible for this line broadening as the width of the central quadrupolar line is inversely proportional to the applied field. 
This holds also for the quadrupolar broadening due to lattice defects. 

\subsection{Splitting of the tetrahedral line}
\label{exchange}

We have demonstrated that not only the disordered \A{} compounds with fcc structure exhibit T--\tp{} splitting in the low-temperature orientationally ordered phase, but also the A$_2$A$\!^\prime$\C{} compounds with sc structure. 
Phenomenologically the splitting is very similar in the two 
structures. 
One notable difference is that in the fcc case the spectral weight of the \tp{} line at low temperatures is about 10{\%} whereas in the sc 
compounds we find a spectral weight of about 30{\%}. 
In this section we address the origin of the T--\tp{} splitting. 

\subsubsection{Sodium site exchange due to reorientations of neighboring \C{} molecules}

The similar temperature dependencies of the spectral weight of the \tp{} line and the width of the \Carbon{} line in \NaCs{} suggest a role of fullerene rotational dynamics in the T--\tp{} splitting.\cite{matus_amr} 
One possible scenario\cite{walstedt,mehring} is that T and \tp{} sites differ in the orientation of surrounding \C{} molecules relative to the sodium ion. 
In this section we show that all our NMR results are compatible with this picture. 

\subsubsection*{\Na{} spectrum}

 \begin{figure*}
 \includegraphics[width=0.7\textwidth]{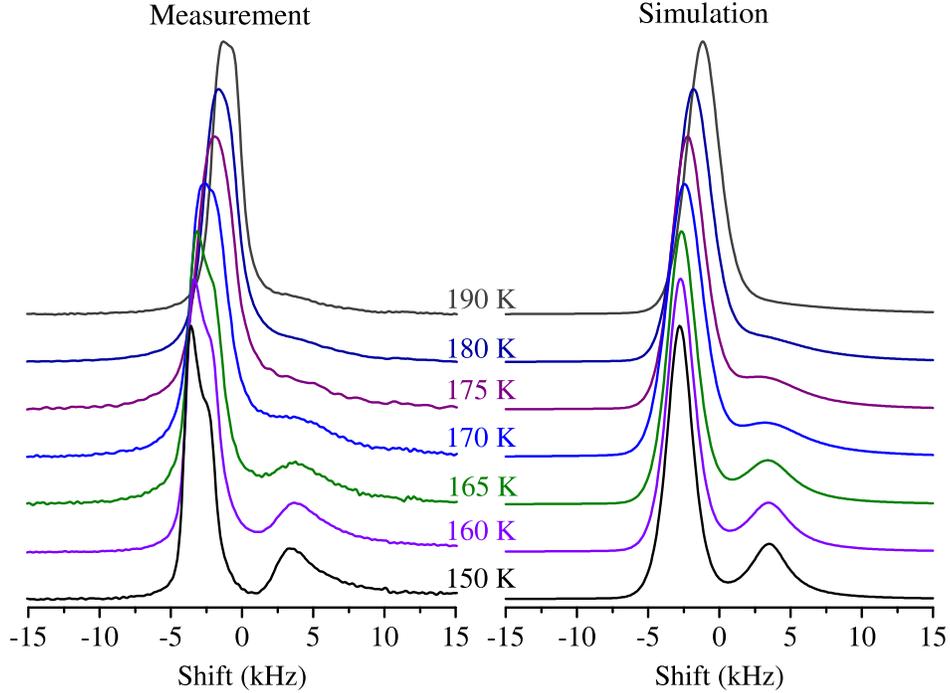}
  \caption{\label{simul} (color online) Measured (left) and simulated (right) sodium NMR spectra of \NaCs{} at several temperatures in the transition region. The simulation is based on the motional narrowing due to site exchange.}
 \end{figure*}

For a quantitative analysis of the influence of fullerene rotational dynamics on the \Na{} spectrum we have performed a line shape simulation. 
The idea is that as the \C{} molecules rotate, a T site becomes \tp{} site and vice 
versa.\cite{mehring} 
If this site exchange is fast enough, the two lines merge into a 
single line, a phenomenon called motional narrowing. 
We describe the process by the following model. 

From the measured low-temperature spectral weights of the two lines $p_1 = 0.71$, $p_2 = 1-p_1 = 0.29$ where $p_1$ and $p_2$ are the probabilities that a given sodium nucleus contributes to the T or \tp{} line, respectively. 
We have obtained these latter values by averaging the measured spectral weights of the two lines, displayed in Fig.~\ref{linewidth}, in the temperature range 10 to 120~K, and the errors represent the scattering in the data. 
If the exchange rate from the T to the \tp{} site is $k_{12}$ and $k_{21} $ is the rate of the inverse process, than the equation of detailed balance reads 

\begin{equation}
\label{balance}
 p_1k_{12}=p_2k_{21} \text{.}
\end{equation}

\noindent
We assume a thermally activated exchange

\begin{equation}
\label{activated}
k_{12} = k_0\exp(-E_a/k_\mathrm{B}T)	
\end{equation}

\noindent
where $E_a$ is the activation energy of the process, $k_0$ is the attempt rate, $T$ is the temperature and $k_\mathrm{B}$ is the Boltzmann constant.  Following the 
treatment of site exchange in standard NMR texts,\cite{abragam,slichter} one can write the 
combined line shape of the two lines as 
 
\begin{equation}
	\label{exchange_spectrum}
	I(\omega)=\Re\, \frac{p_1\left[I_2^{-1}(\omega)+k_{21}\right]+p_2\left[I_1^{-1}(\omega)+k_{12}\right]}{I_1^{-1}(\omega)I_2^{-1}(\omega)-k_{12}k_{21}} \text{.}
\end{equation}

\noindent
Here and in the following subscripts 1 and 2 refer to the T and \tp{} sites, respectively. 
$\Re$ represents real part;

\begin{subequations}
\begin{eqnarray}  
  I_1(\omega)&=&\frac{1}{(\tau_1^{-1}+k_{12})+i(\omega-\delta_1)}  \\
	I_2(\omega)&=&\frac{1}{(\tau_2^{-1}+k_{21})+i(\omega-\delta_2)} 
\end{eqnarray}
\end{subequations}
where $\delta _1$ and $\delta _2$ are the angular Larmor frequencies, 
$\tau_1^{-1}$ and $\tau_2^{-1}$ are Lorentzian broadening parameters, and $i$ the imaginary unit.

For simplicity the quadrupolar broadening of the T line is taken into account with an additional Gaussian broadening of 1.32~kHz. 
A good simultaneous fit is obtained to several spectra taken at different 
temperatures in the vicinity of the line splitting (Fig.~\ref{simul}) with the model parameters assumed to be independent of temperature. 
The best fit parameters 
are $\tau_1^{-1} =2\pi \times 1.3$~kHz, $\tau_2^{-1} =2\pi \times 1.45$~kHz, $E_a 
= 3000\pm400$~K (260~meV), and $k_0 =5\times 10^{10}$ s$^{-1}$. 
In this model the line splitting occurs when the exchange rate becomes slower than the frequency splitting of the lines, $k_{12}\lesssim \Delta\omega_{\mathrm{T\!-T^\prime}}$. 
The activation energies for the site exchange inferred from the temperature dependences of various NMR properties are discussed in Section~B.2.

\subsubsection*{Spin-spin relaxation}

The spin-spin relaxation rate $1/T_2$ can also be described in the framework of site exchange. 
The relaxation rate of spins $I=3/2$ can be written as\cite{mehring_book,mehring}

\begin{equation}
	\label{t2quad}
	\frac{1}{T_2}=\frac{1}{10}\Delta\omega_M^2 J(\Delta\omega_M)+\frac{27}{200}\Delta\omega_Q^2
	J(\Delta\omega_Q)+\frac{1}{T_{2\text{ cst}}}
\end{equation}

\noindent
where the first two terms represent the contribution of local field fluctuations and $T_{2\;\mbox{cst}}$ is a temperature-independent constant representing dipole--dipole and other temperature independent processes. 
$J(\Delta \omega_M )$ and $J(\Delta \omega_Q )$ are the spectral amplitude densities of the magnetic and quadrupolar fluctuations, respectively. 
For the density function we use a Lorentzian form

\begin{equation}
	\label{fluct_dens}
	J(x)=\frac{2\tau}{1+(x\tau)^2}
\end{equation}

\noindent
where $\tau$ is the correlation time of the fluctuations. We take $\tau=k_{12}^{-1}$ and assume again a thermally activated site exchange. 
From the best fit of this model to the measured spin-spin relaxation rate of the T line we obtain $E_a= 2900 \pm 200$~K (250~meV), $k_0=2\times 10^{11}$~s$^{-1}$, $\Delta \omega _M$ = 2$\pi \times$ 2500~s$^{-1}$, and $\Delta \omega_Q$ = 2$\pi \times$ 500~s$^{-1}$. 
The local field fluctuation amplitudes are in the order of the T--\tp{} line splitting frequency indicating the consistency of the site-exchange approach. 
 
\subsubsection*{Spin-lattice relaxation}

The spin-lattice relaxation rate $R\! =\! 1/T_1$ is strongly temperature dependent in the whole range investigated. A common rate for the two tetrahedral sites is observed above 125~K. In the high-temperature range above about 210~K but below the sc-fcc phase transition the relaxation is well described by thermal activation

\begin{equation}
	R_\text{HT}=R_0 \exp (-E_a/k_\text{B}T)
	\label{rht}
\end{equation}

\noindent
as expected for site exchange. 
From a fit to the relaxation rate between 210~K and 270~K, we infer an activation energy $E_a = 3200 \pm100$~K (280~meV) and a prefactor related to the attempt frequency $R_0=3\times 10^7$~s${}^{-1}$ (see the inset of Fig.~\ref{t1}). 
In the sc-fcc phase transition the relaxation rate drops by a factor of 9 and decreases with increasing temperature above the transition. This behavior is seen at the orientational ordering transition in a variety of \C{} systems because in the high-temperature nearly free-rotation state the molecular reorientations are too fast to be efficient in causing spin-lattice relaxation and become less efficient with increasing temperature, \emph{i.e.}, faster rotation. 

Below about 210~K the temperature dependence of $R$ is weaker than exponential. 
In our interpretation at this temperature the Korringa relaxation becomes dominant over the site exchange mechanism as discussed in more details below. 

It is a striking feature of the relaxation rate with far-reaching consequences that even below the temperature of line splitting (170~K) the spin-lattice relaxation rates of the T and \tp{} sites continue to be identical down to 125~K where the relaxation rate curves for the two sites bifurcate. 
The phenomenon is well understood\cite{brouet2} in the framework of the site exchange model. 
We propose that the line splitting temperature and the bifurcation temperature of $R$ are different because of the different time scales involved. 
The fullerene molecular reorientations become slow on the time scale of the inverse of the line splitting frequency but stay fast on the longer time scale of $T_1$ down to the bifurcation temperature.
To model the effect of site exchange on the spin-lattice relaxation rate we use the following rate equations: 

\begin{eqnarray}
	\frac{dM_1}{dt}&=& -(R_1+k_{12})M_1+k_{21}M_2 \label{m1}\\
	\frac{dM_2}{dt}&=& -(R_2+k_{21})M_2+k_{12}M_1 \label{m2}
\end{eqnarray}

\noindent
where the $M_i$'s are the longitudinal magnetizations of the respective sites and the $R_i$'s are the relaxation rates in the absence of site exchange. The $k_{ij}$'s are the site exchange described by Eqs.~(\ref{balance}) and (\ref{activated}).

\begin{table*}
	\caption{\label{tblactivated} Exchange parameters: Activation energies and attempt rates of the site exchange inferred from various experiments.}
	\begin{ruledtabular}
	\begin{tabular}{lccc}
		Experiment& Temperature  & Activation energy & Attempt rate \\
			&					range (K)  &  $E_a$ (K) & $k_0$ (s$^{-1}$)\\
  	\hline
  	Motional narrowing & 150--200 & $3000\pm400$ & $5\times 10^{10}$ \\
  	$T_2$ & 125--225& $2900\pm200$ & $2\times 10^{11}$\\
  	$T_1$ & 125--299 & $3450\pm100$ & $1.5\times 10^{12}$ \\
  	Inelastic neutron scattering\footnotemark[1] & 100--350 & 3500 & $3.9\times 10^{12}$\footnotemark[1] \\
	\end{tabular}
	\end{ruledtabular}
	\footnotetext[1]{From Ref.~\onlinecite{christides}; libron frequency is indicated in column 4.}
\end{table*}

In the slow-exchange limit $k_{12}\! \ll\! R_1,\,R_2\,$ Eqs.~(\ref{m1}) and (\ref{m2}) reproduce the individual relaxation rates of the two sites without exchange, $R_1$ and $R_2$, whereas in the fast exchange limit $k_{12}\! \gg\! R_1,\, R_2$ both sodium sites have a common relaxation rate $R\! =\! p_1R_1\! +\! (1\!-\! p_1)R_2$ where $p_1$ is the spectral weight of component 1.
To solve Eqs.~(\ref{m1})--(\ref{m2}), we use the following form for the relaxation rates: 

\begin{eqnarray}
	R_1&=&A_1T+BT^3+R_\text{HT} \label{t1rate}\\
	R_2&=&A_2T+BT^3+R_\text{HT} \nonumber
\end{eqnarray}

\noindent

The first term describes the Korringa relaxation and the last term corresponds to molecular reorientations as discussed above. The origin of the anomalous $T^3$ term is discussed in Section~\ref{libration}. 
As indicated in Fig.~\ref{t1}, a good fit to the measured relaxation rates is obtained with the following parameters: $A_1=2\times 10^{-5}$ s${}^{-1}$K${}^{-1}$, $A_2=1.3\times 10^{-3}$ s${}^{-1}$K${}^{-1}$, $B=1.85\times 10^{-5}$ s${}^{-3}$K${}^{-1}$, $R_0=3\times 10^{7}$~s${}^{-1}$, $k_0=1.5\times 10^{12}$ s${}^{-1}$, and $E_a=3450\pm100$~K. 

\subsubsection{Dynamical crossover} 
\label{crossover}

Several authors have proposed that the T--\tp{} splitting may result from a phase transition or the separation of two phases.\cite{yoshke} 
Based on our NMR results, we argue in this section against these scenarios and propose that the origin of line splitting is a dynamical crossover. 

The most reliable test for phase separation is SEDOR. Just as in \A{} systems of fcc structure, a phase separation at the temperature of the line splitting in \NaCs{} is excluded on the basis of our \Na{} SEDOR results.\cite{matus_jsc} 

An argument in favor of a phase transition at the splitting temperature is the peak in the spin-spin relaxation rate at the splitting. 
A weak anomaly in the order of experimental error in the differential thermal analysis of \K{} has also been reported.\cite{yoshke} 
As we demonstrate below, our NMR data in \NaCs{} are inconsistent with a phase transition and highly suggestive of a gradual dynamical crossover in the frequency of site exchange. 
The essential signature of a dynamical crossover is that characteristic features of the measured material properties occur at different temperatures depending on the characteristic time scale of the measurement. 

In our case the peak in $1/T_2$ and the line splitting coincide whereas the splitting in $1/T_1$ occurs at a different temperature. 
However, the coincidence of the first two features is evident in the dynamical crossover picture since the characteristic time is the inverse of the frequency splitting of the lines in both cases.
Conversely, the splitting of the $1/T_1$ curves occurs at a markedly different temperature which is inconsistent with the phase transition scenario but consistent with a dynamical crossover as the characteristic time is $T_1$ in this latter case. 
Another argument against a phase transition is the temperature dependence of the average line shift (Fig.~\ref{shift}); no anomaly in this quantity is observed at the line splitting temperature. 

In the previous subsections we have shown that the temperature variation of $T_1$ and $T_2$ as well as the splitting of the \Na{} line are well described assuming a temperature-activated site exchange between the T and \tp{} sites. 
In Table~\ref{tblactivated} we summarize the activation energies inferred from these various experiments together with the temperature range of data from which the activation energy has been obtained. 
The site exchange can be characterized with the same activation energy $E_a = 3300 \pm 240$~K in the whole temperature range 125 to 300~K. 
The activation energy can also be estimated from the energy of the \C{} librational modes measured with inelastic neutron scattering.\cite{cristofolini,christides} 
Assuming that the potential as a function of the angular deflection of the molecule is sinusoidal, an activation energy of 300~meV (3500~K) is obtained from the measured libron energy 2.7~meV. 
This rough estimate agrees well with the NMR results. 
The attempt rate of the site exchange, $k_0$ in Eq.~(\ref{activated}) is also indicated in Table~\ref{tblactivated}. 
Given the exponential temperature dependence, the values obtained for the prefactor from the different experiments agree very well with each other and with the measured libron frequency. 
The temperature-independent parameters of the site exchange are in agreement with the inelastic neutron scattering result of temperature-independent libron energy in the range of 100 to 300~K. 

\subsection{Origin of line splitting in \NaA{}}

\begin{table*}
\caption{\label{tblconfig} Calculated probabilities and second moments as well as building blocks of local fullerene configurations surrounding the tetrahedral site.}
\begin{ruledtabular}
\begin{tabular}{cccccccc}
 \multicolumn{3}{c}{} &\multicolumn{3}{c}{Tetrahedral ions facing} & \multicolumn{2}{c}{No. of C$_{60}$'s orientation} \\
Name & Probability & $\Delta^2$~(ms$^{-2}$) & hexagons & double bonds &  pentagons & major & minor \\
\hline \\
A & 0.689 & 0.1518 & 1 & 3 & 0 & $\left\{\genfrac{}{}{0pt}{0}{4}{3}\right.$\hspace{2mm} & $\genfrac{}{}{0pt}{0}{0}{1}$  \\
B & 0.274 & 0.1525 & 1 & 2 & 1 & $\left\{\genfrac{}{}{0pt}{0}{3}{2}\right.$\hspace{2mm} & $\genfrac{}{}{0pt}{0}{1}{2}$  \\
C & 0.036 & 0.1532 & 1 & 1 & 2 & $\left\{\genfrac{}{}{0pt}{0}{2}{1}\right.$\hspace{2mm} & $\genfrac{}{}{0pt}{0}{2}{3}$  \\
D & 0.001 & 0.1538 & 1 & 0 & 3 & $\left\{\genfrac{}{}{0pt}{0}{1}{0}\right.$\hspace{2mm} & $\genfrac{}{}{0pt}{0}{3}{4}$  \\
\end{tabular}
\end{ruledtabular}
\end{table*}

There is little doubt that just below the fullerene orientational ordering transition the narrow \Carbon{} line and both the \Carbon{} and \Na{} spin-lattice relaxation result from the rapid reorientation of the fullerene molecules. 
This phenomenon has been thoroughly analyzed in a broad range of \C{} compounds based of a wealth of experimental techniques. 
Since we observe a single activation energy for the \Na{} site exchange down to 125~K, it follows that the site exchange process is due to fullerene reorientation. 

Another argument is the concomitant \Na{} line splitting and \C{} line broadening at 170~K.\cite{matus_amr} This behavior also suggests that the \Na{} site exchange is the consequence of different fullerene orientational environments of the T and \tp{} sites. 
In the following we analyze quantitatively how the NMR results can be explained by different fullerene orientational environments. A similar approach has been applied to \Rb{} by Kraus \emph{et al.}\cite{luders}

The orientation of fullerene molecules in \NaCs{} has been studied with x-ray\cite{prassides_xray} and neutron diffraction\cite{prassides} and found to be very similar to pure \C{} (Ref.~\onlinecite{david}).
We recall that in the orientationally ordered \Pa{} structure the molecules are rotated away from the standard orientation by 98$^\circ$ around the threefold symmetry axis through the center of the molecule. 
The typical orientational defect corresponds to an angle of rotation of 38$^\circ$ around the same axis.  The energy difference between these two orientations is very small, in the order of 10~meV (Ref.~\onlinecite{launois}). 

The fraction of fullerene molecules in the majority orientation, $p_{98}$, is found in neutron scattering experiments to increase with decreasing temperature in agreement with thermodynamic expectations. 
In the $T\!\rightarrow\! 0$ limit, however, $p_{98}$ tends to about 0.88, \emph{i.e.}, a value different from 1. 
A similar situation is encountered in pure \C{} where a marked glass transition is observed at $T\! \approx$ 90~K by neutron scattering\cite{david} and thermodynamical measurements.\cite{c60_heatcapacity} 
In \NaCs{} $p_{98}$ varies little below about 140~K without any sharp feature.\cite{prassides_priv} 
The successful description of our $T_1$ data with the same activation energy as at higher temperatures suggests that the glass transition is below 125~K. 

 \begin{figure}
  \includegraphics[width=0.7\columnwidth]{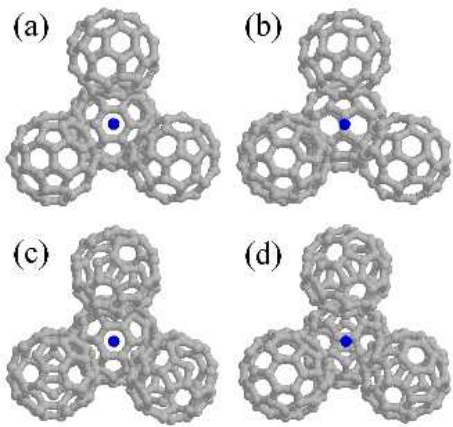}
  \caption{\label{localorder} (color online) \C{} local order around the tetrahedral interstices seen from Na$^+$ ions. If all neighboring fullerene molecules take the majority orientation, the sodium ion faces (a) one hexagon and (b) three double bonds. In case of all neighboring fullerene molecules are in the minority orientation Na$^+$ faces (c) one hexagon and (d) three off-centered pentagons.}
 \end{figure}

Next we consider what local environments of the sodium atoms are formed by its 4 nearest neighbor fullerene molecules (see Fig.~\ref{localorder}). 
We only consider the two frequent orientations.
The sodium atom sits on the threefold symmetry axis of exactly one of its \C{} neighbors. 
This axis goes true the center of a \C{} hexagon in both orientations. 
The other three neighbors turn a C--\,C double bond towards the sodium atom in the majority (98${}^\circ$) orientation and an off-centered pentagon in the minority (38${}^\circ$) orientation. 
The possible fullerene environments are listed in Table~\ref{tblconfig}. If we assume that the two orientations are spatially uncorrelated and use $p_{98} = 0.88$, the neutron diffraction result\cite{prassides} in the low-temperature limit, the frequency of these configurations are readily calculated; the results are listed in Table~\ref{tblconfig}.

The different local environments should result in different \Na{} NMR frequencies with spectral weights equal to the frequencies of the corresponding configurations. 
The frequencies of the two most frequent configurations, labeled A and B in Table~\ref{tblconfig}, are $p_A = 0.69$ and $p_B = 0.27$. 
These values are very close to the measured NMR spectral weights of the T and \tp{} lines in the low-temperature limit, $0.71 \pm 0.05$ and $0.29 \pm 0.05$, respectively. 

The remaining two configurations, C and D in Table~\ref{tblconfig}, carry a combined spectral weight of 5\% and are easily missed in the NMR measurements with the resolution we have, especially if these lines are broad or overlap with the two observed lines.

\subsection{The influence of librations on $\boldmath{T_1}$ relaxation rates}
\label{libration}

In this section we address the origin of the surprising $T^3$ term in the $1/T_1$ relaxation rate (see Eq.~(\ref{t1rate}) and Fig.~\ref{t1}). 
In the discussion of the spin-lattice relaxation we conclude that below about 200~K, the relaxation rate is governed by the local electronic susceptibility, rather than by molecular motion. 
In this situation the Korringa relation is expected to hold:

\begin{equation}
\label{libr}
 K^2=S\ \frac{1}{T_1T}\text{,}
\end{equation}

\noindent
where $S$ is the temperature independent Korringa constant. 
Indeed, the behavior of the Knight shift is also anomalous. 
Clearly the Knight shift has a contribution linear in temperature, although signs of saturation are observable in the low-temperature limit. 
A linear contribution in the Knight shift implies the existence of a cubic term in the relaxation rate as seen in our experiments. 

 \begin{figure}
  \includegraphics[width=0.95\columnwidth]{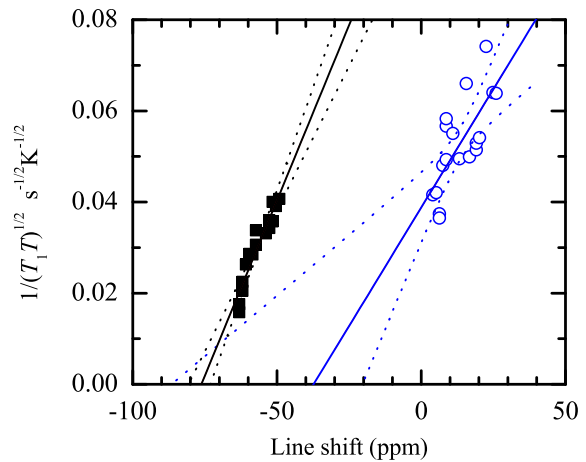}
  \caption{\label{sqkorr} (color online) $1/(T_1T)^{1/2}$  s a function of line shift. The solid lines are best linear fits and the dotted lines are 95\% confidence bands.}
 \end{figure}

To test the validity of the Korringa relation, in Fig.~\ref{sqkorr} we plot $1/(T_1T)^{1/2}$ against $K$ for both the T and \tp{} lines together with the best linear fits. 
A linear dependence is well obeyed for the T line. 
Since the intensity of the \tp{} line is smaller, the experimental errors are larger for this line both in the $T_1$ and $K$ data. 
By extrapolating the linear relation to $1/(T_1T)^{1/2}=0$, the zero point of the Knight shift, \emph{i.e.}, the chemical shift is obtained. 
The chemical shift for the T line is $-76 \pm 5$~ppm. 
Because of the substantial experimental error in the \tp{} data, the 95\% statistical confidence bands of the linear fit are also included in the figure. The chemical shift for the \tp{} line is $-37$~ppm with a 95\% confidence interval of $-87$ to $-20$~ppm. 

From the slope of the linear fit we obtain the Korringa constant $S = (4.2 \pm 0.5) \times 10^{-7}$~sK and $(9.3 \pm 1.5) \times 10^{-7}$~sK for the T and \tp{} lines, respectively. These values are extremely small compared to the theoretical value for the isotropic contact interaction between the nuclear and electronic spins, $S = 3.8 \times 10^{-6}$~sK. 
This finding indicates that the anisotropic magnetic dipole-dipole interaction dominates the coupling of the nuclear and electronic spins. 
Such a coupling can only arise from the interaction of the nuclei with the oriented carbon $p_z$ orbitals.

Next we ask what may cause the temperature dependence of the \Na{} Knight shift. 
One straightforward possibility is the existence of strong electronic correlations. 
Electronic correlations may lead to a temperature-dependent enhancement of the electronic susceptibility. However, the \Carbon{} and $^{133}$Cs relaxation rates, and, most importantly, the EPR susceptibility are all temperature independent\cite{brouet3} below 100 K. The same observation excludes thermal expansion as the origin of the temperature dependent \Na{} relaxation rate. What can be then the low-lying excitation responsible for the anomalous Knight shift and relaxation rate? 
In \C{} compounds including \NaCs{} there exists a low-lying optical mode involving the rigid librational motion of \C{} molecules. 
The energy of librons in \NaCs{} is found to be $\hbar \omega_0/k_\text{B} =31$~K (2.7~meV) by inelastic neutron scattering.\cite{cristofolini} The result on the Korringa constant suggest that the exact orientation of the carbon $p_z$ orbitals has a strong influence on the \Na{} Knight shift. 
Therefore the librons may strongly influence the Knight shift. 
Up to second order in the angular excursion of the \C{} molecules, the Knight shift is written as

\begin{eqnarray}
\label{knight2}
K & = & K_0 + \widetilde{K}_1 \langle A  \cos \omega_0 t\rangle + 2\widetilde{K}_2 \langle  A^2 \cos^2 \omega_0 t\rangle = \nonumber \\  
& & K_0 + \widetilde{K}_2 \langle A_2 \rangle \text{, }
\end{eqnarray}

\noindent
where $K_0$ is the Knight shift in the absence of librons, $\omega_0$ is the libron frequency, $A$ is the libron amplitude, $\widetilde{K}_1$ and $\widetilde{K}_2$ are coefficients that depend on the shape of the carbon orbitals as well as on the relative orientation of the \C{} molecules with respect to the sodium nucleus. 
A time average---indicated by angular brackets---has to be taken since $\omega_0 \gg \omega_L$. 

For $k_\text{B} T \gg \hbar \omega_0$, $\langle A^2 \rangle \propto T$ yielding a contribution to the Knight shift linear in $T$. 
For a stricter condition on the lower bound of $T$-linear behavior, we take the Einstein model of lattice vibrations in which the temperature-dependent part of the amplitude square is given by

\begin{equation}
\label{einstein}
\langle A^2 \rangle \propto \hbar \omega_0/[(\exp (\hbar \omega_0/k_\text{B}T) - 1)]\text{.}
\end{equation}

\noindent
$\langle A^2 \rangle$ in this model is linear in $T$ with good approximation down to $\omega_0/2$ (15~K), \emph{i.e.}, to the lowest temperature we have investigated. This is demonstrated in Fig.~\ref{shift} where a fit of the form 

\begin{equation}
\label{einsteinfit}
K=K_0 + K_2\, \hbar \omega_0/k_\text{B}/ (\exp( \hbar\omega_0/k_\text{B}T) -1 )
\end{equation}

\noindent
to the measured Knight shift is given. The fit parameters are $K_2 = 0.17\times10^{-6}$~K$^{-1}$ and $0.27\times 10^{-6}$~K$^{-1}$ for the T and \tp{} lines, respectively. 
Since the local \C{} environments are different at the T and \tp{} sites, the different fit parameters for the two lines are natural.

This interpretation is in agreement both with the anomalous value of the \Na{} Korringa constant and with the different temperature dependence of the \Na{} and \Carbon{} spin-lattice relaxation rates since the latter nucleus is expected to be much less sensitive to the angular orientation of the \C{} molecules because the conduction band is formed of carbon orbitals. 
Remarkably, there is a big jump in \Na{} $1/T_1$ at the orientational ordering transition (see Fig.~\ref{t1}) while the jump is unresolved within experimental error in \Carbon{} $1/T_1$ (Ref.~\onlinecite{brouet3}).
The jump in the \Na{} case is well understood with the assumption that the excess Knight shift is proportional to the mean-square libron amplitude. 
The difference between the Knight shift in the fcc phase and in the low temperature limit is 190~ppm, whereas the excess Knight shift acquired in the sc phase until the phase transition is reached is 60~ppm. 
If one considers a ``Lindemann criterion of orientational melting,'' the cited data correspond to a Lindemann constant $c_L = \sqrt{60/190} = 0.56$, a realistic number. 
The jump in the lattice constant at the phase transition, on the other hand, is only 25\% of the increase acquired in the sc phase. 
Assuming that the excess Knight shift is the result of the increase in the DOS due to thermal expansion, the big jump as well as the different behaviour of the \Carbon{} $1/T_1$ are difficult to interpret. 

Finally we note that a similarly strong temperature dependence of $^{39}$K $1/(T_1T)$ has been observed in \K{} of fcc structure.\cite{yoshke_unpublished} 
It remains to be seen if the phenomenon can be interpreted in terms of \C{} librations in this material of different orientational structure.

\section{Summary}

We have demonstrated in a series of alkali-fulleride superconductors of simple cubic structure, \emph{i.e.} in  \NaCs{}, \NaRb{}, and \NaK{}, that the NMR line associated with the alkali nucleus of tetrahedral \C{} coordination splits into two lines at low temperature. 
The analysis of the spectrum and SEDOR indicates that these two lines originate from the same phase. 
By measuring the temperature dependence of the \Na{} NMR spectrum, spin-lattice and spin-spin relaxation times in \NaCs{}, there is clear evidence that the origin of the splitting of the NMR line is a dynamic two-site exchange. 
The exchange rate is temperature activated in the temperature range 125 to 299~K with a single activation energy of 3300~K. 
This finding rules out the existence of a phase transition in the above temperature range. 
We propose that the two sites differ in the fullerene orientational environment. 
The different environment reflect the two different \C{} orientations observed by neutron diffraction.
We find that the Knight shift and $1/(T_1T)$ are temperature dependent down to 10~K. We interpret this result in terms of \C{} librational modes.

\begin{acknowledgments}

We are indebted to thank for the many fruitful discussions with R.~Moret, K.~Prassides, G.~ Oszl{\'a}nyi, G.~Zar{\'a}nd, K.~Tompa and F.~I.~B.~Williams. Research in France has been supported by EU Marie Curie Training Site in ``STROCOLODI'' program under the contract HPMT-CT-2000-00042, in Switzerland by the Swiss National Science Fundation and its NCCR ``MaNEP'' program and in Hungary by Grant OTKA T0373976. One of the authors (P.M.) would like to express his sincere appreciation for the hospitality of the Orsay group.
\end{acknowledgments}


\end{document}